\documentclass[a4paper,11pt]{article}
\usepackage{jinstpub} 
\usepackage{lineno}
\usepackage{subfig}
\usepackage{comment}
\usepackage{enumitem}

\newcommand{\mycomment}[1]{}{\bf }

\usepackage{float}

\notoctrue

\DeclareOldFontCommand{\rm}{\normalfont\rmfamily}{\mathrm}
\DeclareOldFontCommand{\sf}{\normalfont\sffamily}{\mathsf}
\DeclareOldFontCommand{\tt}{\normalfont\ttfamily}{\mathtt}
\DeclareOldFontCommand{\bf}{\normalfont\bfseries}{\mathbf}

\title{Compact Representation of Particle-Collision Events for Physics-Informed Machine Learning}

\author[a]{W.~Islam,}
\author[b]{S. V. ~Chekanov}

\affiliation[a]{Department of Physics, University of Wisconsin, Madison, WI, USA.}
\affiliation[b]{HEP Division, Argonne National Laboratory, 9700 S.Cass, Lemont, IL 60439, USA}

\emailAdd{wasikul.islam@cern.ch, chekanov@anl.gov}

\abstract{
\begin{abstract}

We introduce a compact, physics-driven event representation, \emph{RMM-C46}, designed to compress the high-dimensional rapidity–mass matrix (RMM) into a low-dimensional, interpretable feature set suitable for physics-informed machine-learning (ML) and quantum-computing applications. The full RMM encodes detailed pairwise correlations among jets, $b$-jets, leptons, photons, and missing transverse energy but contains more than a thousand values per event, making it computationally heavy for large-scale training and incompatible with current low-qubit quantum devices. The proposed RMM-C46 input space for ML preserves the physical block structure of the RMM through aggregated invariant-mass, rapidity-difference, and transverse-energy components, reducing sizes of the original RMM by over an order of magnitude while maintaining interpretability. Applied to simulated proton–proton collisions at $\sqrt{s}=13.6~\mathrm{TeV}$, these representations match or exceed the discriminative performance of the full RMM in both supervised and unsupervised ML tasks. Their compactness, stability, and physics transparency also make them naturally compatible with near-term quantum machine-learning architectures.
RMM-C46 provides scalable, efficient, and quantum-ready alternatives to the full RMM for next-generation collider-physics analyses.
\end{abstract}
}


\arxivnumber{} 

\note{Preprint: ANL-HEP- 202401}

\begin{document}

\maketitle
\flushbottom

\clearpage

\section{Introduction}
\label{sec:intro}

The search for new physics in high-energy proton--proton collisions increasingly relies on machine-learning (ML) methods capable of identifying subtle deviations from Standard Model (SM) expectations.  
Unsupervised anomaly detection has become a promising paradigm for model-agnostic searches~\cite{ATLAS:2023ixc,BELIS2024100091}, enabling sensitivity to unexpected signatures without requiring explicit signal labels or assumptions about new-physics hypotheses.  
However, many existing ML approaches operate directly on high-dimensional or sparsely populated input spaces that do not explicitly encode the physics structure of collider events.  

Recent advances in collider-ML have explored alternative event representations that aim to incorporate physical symmetries or reduce dimensionality at the architecture level. Set-based approaches such as Energy Flow Networks (EFNs) and Particle Flow Networks (PFNs)~\cite{Komiske:2018cqr} exploit permutation invariance in particle collections to construct infrared- and collinear-safe observables directly from low-level inputs. Other studies have introduced Lorentz-equivariant neural architectures~\cite{Bogatskiy:2020tje} that explicitly respect relativistic symmetries of the underlying kinematics, as well as representation-streamlining techniques such as moment pooling~\cite{Gambhir:2024dtf} to reduce redundancy in high-dimensional collider inputs for downstream ML tasks. While these methods offer powerful data-driven representations, they typically rely on learned embeddings that do not retain an explicit mapping to interpretable physics observables.

Such representations can struggle to exploit event topology efficiently and often incur substantial computational overhead during training and deployment.

The rapidity--mass matrix (RMM) formalism~\cite{Chekanov:2018nuh,universe7010019} (see the definition in Sect.~\ref{sec:RMM_definition}) provides a physics-grounded alternative: a structured matrix encoding Lorentz-invariant pairwise correlations among reconstructed jets, $b$-jets, leptons, photons, and missing transverse energy.  
The full RMM has proven highly effective for supervised~\cite{Chekanov:2025xpk} and unsupervised~\cite{ATLAS:2023ixc} collider-ML tasks, but its dimensionality (typically $\sim$1287 entries) poses challenges for large-scale training, introduces redundancy, and limits compatibility with quantum machine-learning architectures where qubit counts remain severely restricted.

One major feature of the RMM is its fixed size and the sparse structure: many entries are zeros used as padding for missing information about objects. The matrix must be large enough to accommodate any LHC event, regardless of how many objects are present, which inevitably introduces these zeros. Although sparsity itself is not problematic, anomaly detection with autoencoders requires careful handling of padding zeros. In particular, it is recommended to mask cells with zero values for all events to prevent the reconstruction loss from being dominated by uninformative entries.

Previous studies \cite{ATLAS:2023ixc} found no clear advantage of variational autoencoders (VAEs) over standard autoencoders when applied to RMMs. A likely reason is that, in practice, VAE-based anomaly detection benefits from compact representations with fewer non-zero values, compared to the original RMM. 
When many missing values are encoded as zeros in a domain where zero is a valid numerical value (or at least not explicitly marked as ``missing''), the model is forced to interpret these zeros as genuine data. This, in turn, leads the learned generative distribution to model artifacts of the padding rather than the underlying physics, degrading anomaly-detection performance since  zeros constitute a large fraction of the input. 
In addition, a smaller number of variables representing LHC events 
is beneficial for relatively simple neural networks and requires much less computational power.
These considerations motivate the development of a more compact RMM representation.

Some recent studies attempt to mitigate this dimensionality bottleneck by compressing collider events into latent features via generic autoencoders or by applying machine-learned particle-flow reconstruction pipelines prior to downstream ML or quantum algorithms~\cite{nature2024quantumAD,Mokhtar:2025zqs}.  

While these hybrid pipelines demonstrate feasibility, the resulting latent coordinates lack explicit ties to the physics block structure of the event and may not preserve the semantics of the original topology.

In this work, we develop a compact, physics-driven alternative to the full RMM: the \emph{RMM-C46} representation.  
These formats are constructed directly from the physical content of the RMM using blockwise Frobenius norms, transverse-energy sums, rapidity-difference aggregates, and invariant-mass structures.  
They provide low-dimensional, interpretable feature vectors that retain nearly all of the discriminative power of the original matrix while drastically reducing the memory footprint and training time.  
Most importantly, their dimensionality ($46$) aligns naturally with the practical qubit capacities of near-term quantum devices, enabling direct angle or amplitude-encoding without further preprocessing. This design philosophy provides an alternative to recent quantum-enhanced learning approaches based on learned graph embeddings or equivariant quantum networks~\cite{Jahin:2024zss}, which typically operate in learned latent or graph-based embedding spaces without preserving the explicit block structure of predefined physics observables in the event representation.

Using Monte Carlo (MC) simulations of $pp$ collisions at $\sqrt{s}=13.6\,\text{TeV}$, we evaluate the performance of these compressed RMM formats on benchmark processes involving heavy scalar resonances decaying into multi-Higgs final states.  
In particular, we study exotic cascades of the form $X\!\rightarrow SH\!\rightarrow HHH$, with $m_S=280~\text{GeV}$ and resonance masses $m_X=0.5$--$2.0~\text{TeV}$, along with direct di-Higgs processes $X\!\rightarrow HH$ producing $\ell\ell b\bar{b}+A$ final states.  
Inclusive $t\bar{t}$ events serve as the dominant SM background, with $WZ$+jets used for cross-validation.  
All samples are generated using \textsc{Pythia8}~\cite{Sjostrand:2007gs}, with signal cross sections obtained from \textsc{MadGraph5\_aMC@NLO}~\cite{Alwall:2014hca} where available, and normalized to $140~\mathrm{fb}^{-1}$.

Our results show that the compressed RMM formats match or exceed the discriminative performance of the full RMM in supervised and unsupervised ML tasks, while providing clearer interpretation and significantly improved computational efficiency.  
By aggregating physically correlated regions of the full matrix into well-defined scalar quantities, the RMM-C46 format establishes a practical and physics-transparent foundation for scalable ML pipelines.  
They also offer one of the first event representations designed explicitly for deployment on low-qubit or hybrid quantum--classical architectures.

The remainder of this paper is organized as follows.  
Section~\ref{sec:RMM-C46_details} defines the RMM-C46 representation and its physics-motivated construction. Section ~\ref{sec:performance} discusses performances of the new formats, and 
section~\ref{sec:quantum_prospects} discusses prospects for classical and quantum ML applications enabled by these compressed representations.

\section{Rapidity--Mass Matrix (RMM)}
\label{sec:RMM_definition}
To facilitate understanding of a compact event representation in the form of a fixed-size data structure, we recall the definition of the rapidity--mass matrix (RMM)~\cite{Chekanov:2018nuh,universe7010019} and rewrite it in an analytical form.

The  RMM is a physics-motivated $m\times m$ array that encodes pairwise correlations among all reconstructed objects in an event.  For RMM, each row/column corresponds to one ``slot'' reserved for a specific object class.
For five types of objects ($T=5$), the following order was typically used:
\text{MET},  jets, $b$-jets, muons, electrons, photons.
For each type, $N$ object slots are reserved (typically $N=10$), so that  
$m = 1 + T\times N = 51$,
giving a $51\times 51$ matrix with $2601$ elements per event. This RMM configuration is called T5N10.
This data structure enables easy visualization of structurally complex kinematic data for many different types of objects.

Below we give the full definition of the RMM. Let $R_{ij}$ denote the element in row $i$, column $j$ of the event RMM. The indices are organized into contiguous blocks as follows:

\[
\begin{aligned}
&\text{MET:}         && i=1, j=1\\
&\text{Jets:}             && i \> \mathrm{or}\>  j \in [2,\;N+1], \\
&\text{$b$-Jets:}         && i \> \mathrm{or}\>  j \in [N+2,\;2N+1], \\
&\text{Muons:}            && i \> \mathrm{or}\>  j \in [2N+2,\;3N+1], \\
&\text{Electrons:}        && i \> \mathrm{or}\>  j\in [3N+2,\;4N+1], \\
&\text{Photons:}          && i \> \mathrm{or}\>  j\in [4N+2,\;5N+1].
\end{aligned}
\]
where MET is scaled by a centre-of-mass energy $\sqrt{s}$. The first row and column  have the following definition:
\[
R_{i1} \ (i>1) = \frac{m_T(t_i)}{\sqrt{s}},
\qquad
R_{1j} \ (j>1) = h_L(t_j).
\]
Here $m_T(t_i)$ is the transverse mass of object $t_i$, and $h_L$ reflects the
longitudinal Lorentz factors~\cite{Chekanov:2018nuh} in the longitudinal direction.
The remaining elements are defined as variables that are Lorentz invariant under boosts along the longitudinal axis:
\[
R_{ij} =
\begin{cases}
\frac{E_T(t_i)}{\sqrt{s}}, & i>1, \> i=j=L,   \\[4pt]
\delta E_T (t_i) \equiv \frac{E_T(t_{i-1}) - E_T(t_{i1})}{E_T(t_{i-1}) + E_T(t_{i})}  , &   i>2, \>  i=j \ne L,\\[4pt]
\dfrac{m(t_i,t_j)}{\sqrt{s}}, & i>j, \\[6pt]
h(t_i,t_j), & i<j.
\end{cases}
\]
where $L$ represents a set of five elements on the matrix diagonal (for $T=5$ types of objects)  defined as $L \in [2, \; N+2, \; 2N+2, \; 3N+2, \; 4N+2]$, reserving the slots for the scaled transverse energies. The other diagonal values are the transverse momentum imbalances,  $\delta E_T (t_i)$.
Finally, $m(t_i,t_j)$ are two-body invariant masses, and $h(t_i,t_j)$ are rapidity differences of two objects. 
Thus, the diagonal elements store transverse energies,   the upper triangle stores pairwise invariant masses,  and the lower triangle stores rapidity differences. The azimuthal angles between objects are not included in the RMM, as they can be inferred from the RMM variables.

This block-structured definition establishes an interpretable geometric map of the event in a single matrix. However, the full $51\times 51$ RMM contains approximately 2,600 entries, many of which are redundant for downstream machine learning. In recent studies \cite{Chekanov:2025xpk, Islam:2025eas}, the number of slots for electromagnetic objects (leptons and photons) has been reduced from 10 to 5 to decrease the number of empty cells, since LHC energies are not sufficient to populate all such slots. Although this reduces the number of variables from 2,601 to 1,287, this data structure still includes many variables that take the value 0 for traditional SM events and the most popular BSM signals.

\section{Compact RMM-C46 Representation}
\label{sec:RMM-C46_details}

In this section, we propose a compressed representation of the ML inputs, RMM-C46, that drastically reduces the number of zero entries while preserving the main kinematic channels of the full RMM matrix.
The number 46 reflects the fact that this representation has 46 zones that have physical motivation.
The starting point is the full RMM $R_{ij}$ defined in Sect.~\ref{sec:RMM_definition}, with indices partitioned
into MET, jets ($j$), $b$-jets ($bj$), muons ($\mu$), electrons ($e$), and photons ($\gamma$).
We then decompose the matrix into 46 disjoint regions, each associated with a specific physical quantity or
pairwise structure:

\begin{figure}[H]
    \centering
    \includegraphics[width=1.0\textwidth]{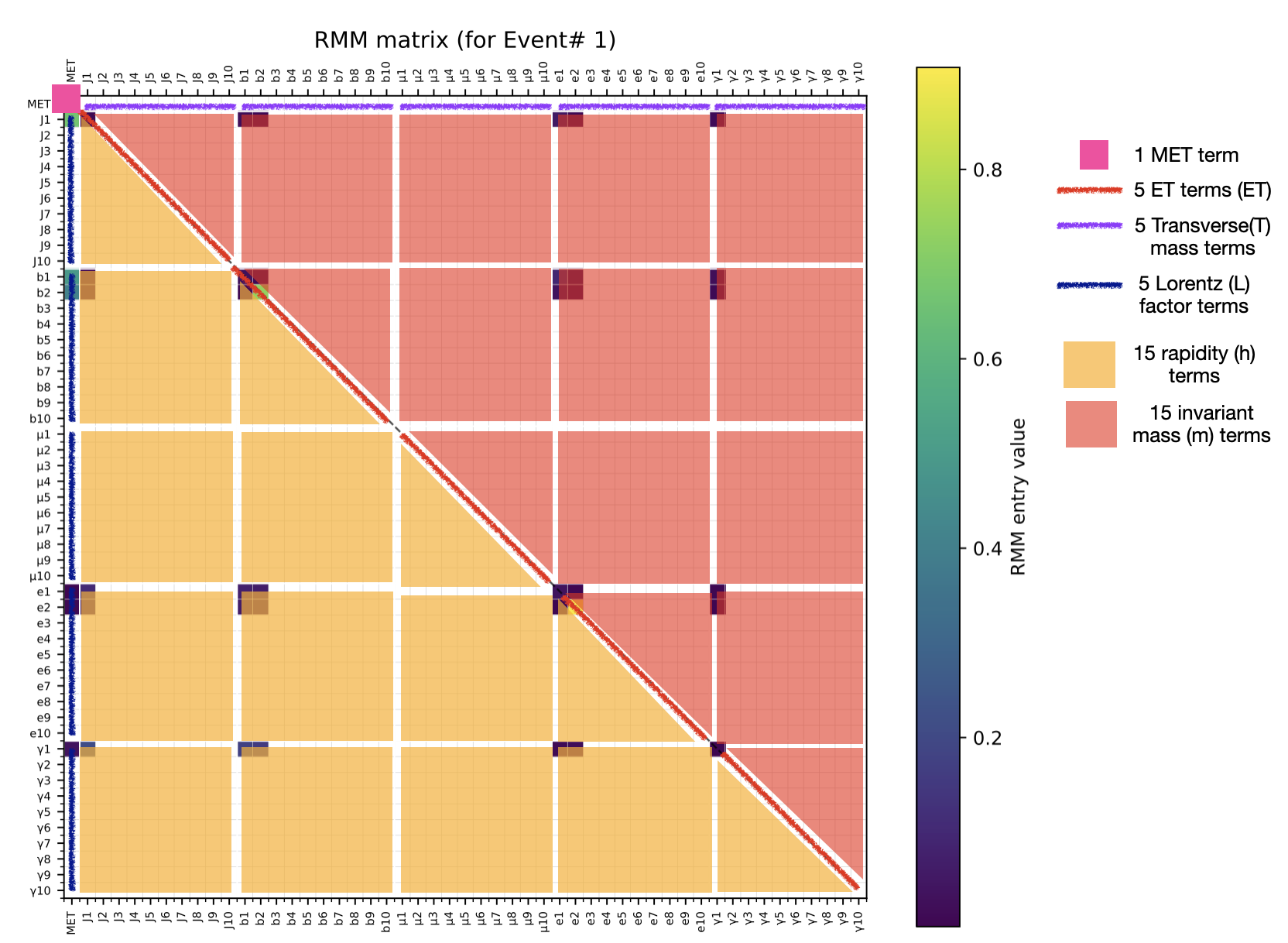}
    \caption{
    Example of a RMM matrix, displaying 46 zones, which result into 46 variables.}
    \label{fig:46_zones}
\end{figure}

\begin{itemize}
  \item 1 global MET term,
  \item 5 transverse-energy terms $E_T$ (one per object class),
  \item 5 transverse-mass--like terms $T$ (one per class),
  \item 5 longitudinal/Lorentz--like terms $L$ (one per class),
  \item 15 rapidity--difference zones $h_{\alpha,\beta}$,
  \item 15 invariant-mass zones $m_{\alpha,\beta}$.
\end{itemize}
The resulting 46-dimensional feature vector is
\[
\mathbf{v}_{\rm RMM\text{-}C46}
=
\big(
  {\rm MET},\> 
  E_T^{(j)},\> E_T^{(bj)},\> E_T^{(\mu)},\> E_T^{(e)},\> E_T^{(\gamma)},\> 
  T^{(j)},\> T^{(bj)},\> T^{(\mu)},\> T^{(e)},\> T^{(\gamma)}, 
\]
\[
  L^{(j)},\> L^{(bj)},\> L^{(\mu)},\> L^{(e)},\> L^{(\gamma)}, 
\]
\[
  h_{j,j},\> h_{bj,j},\> h_{bj,bj},\> h_{\mu,j},\> h_{\mu,bj},\> 
  h_{\mu,\mu},\> h_{e,j},\> h_{e,bj},\> h_{e,\mu},\> h_{e,e},\>  
h_{\gamma,j},\> h_{\gamma,bj},\> h_{\gamma,\mu},\> h_{\gamma,e},\> h_{\gamma,\gamma},
  \]
\[
  m_{j,j},\> m_{bj,j},\> m_{bj,bj},\> m_{\mu,j},\> m_{\mu,bj},\> 
  m_{\mu,\mu},\> m_{e,j},\> m_{e,bj},\> m_{e,\mu},\> m_{e,e},\> 
m_{\gamma,j},\> m_{\gamma,bj},\> m_{\gamma,\mu},\> m_{\gamma,e},\> 
m_{\gamma,\gamma} 
\]
\[
\big)\in\mathbb{R}^{46}.
\]

Each component is computed from a \emph{distinct, non-overlapping} zone of the RMM; the 46 zones form a
partition of the matrix entries used in the construction.
The definitions are as follows:
\begin{itemize}
  \item \textbf{MET term (1 variable).}
  The global MET magnitude is stored in the entry $R_{00}$, and defines a single scalar zone
  \[
    {\rm MET} \;\;\leftrightarrow\;\;  R_{00}.
  \]

  \item \textbf{$E_T$ terms (5 variables).}
  For each object class $t\in\{j,bj,\mu,e,\gamma\}$, we collect the diagonal entries of the corresponding same-type
  block (e.g.\ jets--jets, $b$-jets--$b$-jets, etc.),
  \[
    Z^{(E_T)}_t = \{\, R_{ii} \mid i \in \text{indices of type }t \,\},
  \]
  which encode the transverse energy $E_T$ and its imbalances for that class.

  \item \textbf{$T$ terms (5 variables).}
  For each class $t$, we define a transverse-mass--like zone from the first row of the RMM, restricted to the
  columns of that class (excluding the MET entry),
  \[
    Z^{(T)}_t = \{\, R_{0j} \mid j \in \text{indices of type }t \,\},
  \]
  corresponding to transverse-mass variables $m_T$ built between MET and objects of type $t$.

  \item \textbf{$L$ terms (5 variables).}
  For each class $t$, we define a longitudinal/Lorentz--like zone from the first column, restricted to the rows
  of that class,
  \[
    Z^{(L)}_t = \{\, R_{i0} \mid i \in \text{indices of type }t \,\},
  \]
  which encode longitudinal or rapidity-related kinematics associated with objects of type $t$.

  \item \textbf{Rapidity zones $h_{\alpha,\beta}$ (15 variables).}
  For each ordered pair of classes $(\alpha,\beta)\in\{j,bj,\mu,e,\gamma\}$ we define one rapidity zone.
  For same-type pairs ($\alpha=\beta$) the $h_{\alpha,\alpha}$ zone is the strictly lower triangle of the
  corresponding block,
  \[
    Z^{(h)}_{\alpha,\alpha} = \{\, R_{ij} \mid i,j \in \alpha,\; i>j \,\},
  \]
  which stores pairwise rapidity differences $\Delta y$ within the class.
  For cross-type pairs ($\alpha\neq\beta$), $h_{\alpha,\beta}$ is formed from the dedicated rapidity block
  as indicated by the original RMM layout (e.g.\ $h_{bj,j}$ from the $b$-jet--jet rapidity cells, $h_{e,\mu}$
  from the electron--muon rapidity cells, etc.),
  \[
    Z^{(h)}_{\alpha,\beta} = \{\, R_{ij} \mid (i,j) \in \text{rapidity block for }(\alpha,\beta) \,\}.
  \]

  \item \textbf{Mass zones $m_{\alpha,\beta}$ (15 variables).}
  Analogously, each mass zone collects the invariant-mass entries for a given pair of classes.
  For same-type pairs ($\alpha=\beta$), the $m_{\alpha,\alpha}$ zone is the strictly upper triangle of the
  same-type block,
  \[
    Z^{(m)}_{\alpha,\alpha} = \{\, R_{ij} \mid i,j \in \alpha,\; i<j \,\},
  \]
  where $R_{ij}\propto m_{ij}/\sqrt{s}$ encodes pairwise invariant masses.
  For cross-type pairs ($\alpha\neq\beta$), $m_{\alpha,\beta}$ is defined by the corresponding mass block in the
  RMM (e.g.\ jets--$b$-jets, muons--jets, electrons--photons, etc.),
  \[
    Z^{(m)}_{\alpha,\beta} = \{\, R_{ij} \mid (i,j) \in \text{mass block for }(\alpha,\beta) \,\}.
  \]
\end{itemize}

By construction, the 46 zones
\[
Z_{\rm MET},\; Z^{(E_T)}_t,\; Z^{(T)}_t,\; Z^{(L)}_t,\; Z^{(h)}_{\alpha,\beta},\; Z^{(m)}_{\alpha,\beta}
\]
are disjoint and cover all cells of the RMM that participate in the C46 representation.
What remains is to specify how each zone is aggregated into a single scalar feature. 

\subsection{Zone Aggregation Schemes for RMM-C46}
\label{subsec:RMM-C46_agg}

Each of the 46 RMM-C46 variables is obtained by aggregating the entries of one
physically defined RMM zone.  
Two complementary aggregation rules are used:

\begin{itemize}
\item \textbf{Additive aggregation} (\emph{RMM-C46-add}):  
each feature is the simple sum of all RMM entries in its zone,
\[
v_k^{\rm (add)}=\sum_{(i,j)\in Z_k} R_{ij}.
\]
This yields a signed measure of the total activity in that block and scales
linearly with the number and magnitude of contributing cells.

\item \textbf{Frobenius aggregation} (\emph{RMM-C46-frob}):  
each feature is the Frobenius norm of the zone,
\[
v_k^{\rm (frob)}=\sqrt{\sum_{(i,j)\in Z_k} R_{ij}^2}.
\]
This produces a strictly non-negative “energy-like’’ measure that emphasizes
large entries, since contributions grow quadratically.
\end{itemize}

\begin{figure}[H]
  \begin{center}
    \includegraphics[height=0.95\textheight,width=0.55\textwidth]{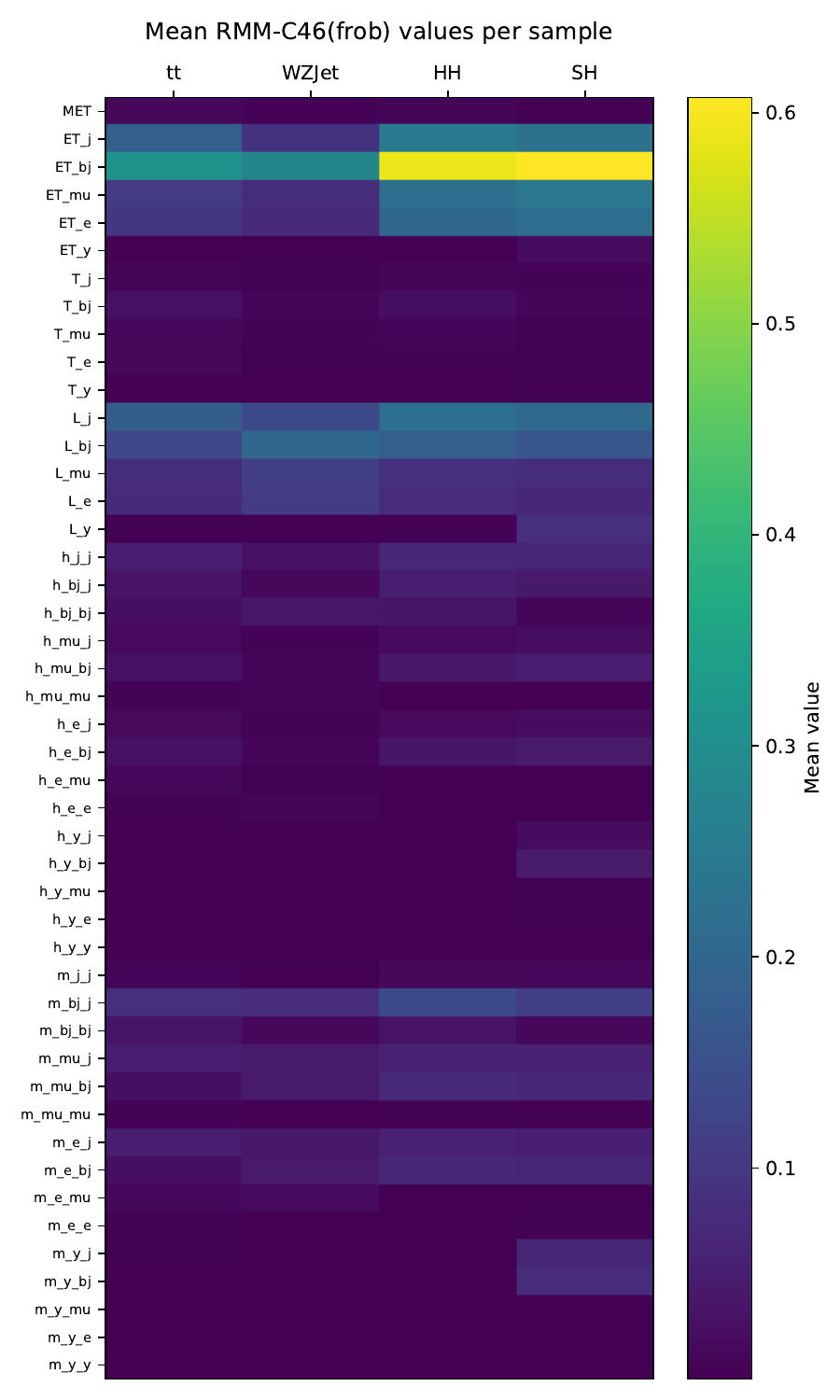}
  \end{center}
\caption{Comparison of the mean RMM-C46 feature values for four datasets:
    Standard Model $t\bar t$, WZJets, $X\!\rightarrow\!HH$ (1500~GeV), 
    and $X\!\rightarrow\!SH$ (1500~GeV).
    The Frobenius aggregation heatmap 
    (\emph{C46-frob}) shows the 46-zone partition of the RMM, and  highlights 
    different aspects of the underlying block magnitudes.}
\label{fig:C46_mean_heatmap_compare}
\end{figure}

Both variants rely on the same 46-zone partition of the rapidity-mass matrix.
The additive scheme preserves the signed and cumulative structure of each block,
while the Frobenius scheme measures its geometric size through a quadratic
aggregation.  
Although both encodings exhibit broadly similar behaviour across the
$t\bar{t}$, WZJets, $X\!\to HH$, and $X\!\to SH$ samples, the Frobenius
aggregation consistently provides sharper contrast in zones dominated by a few
large entries-such as those arising from hard or boosted objects-while the
additive scheme is more sensitive to the number of non-zero cells in a block.

Because the Frobenius variant enhances physically relevant high-energy
structures and yields more stable behaviour across signal masses, we adopt
\emph{RMM-C46-frob} as the default compact representation for the remainder of
this study.

To give some insight about the RMM-C46, let us consider events with only two jets (and no substantial MET). Such events will populate three elements: indices 2, 12, and 17; all other elements will be zero. For events with three (or more) jets, the additional index 32 will also have a non-zero value. High-multiplicity events will not increase the number of non-zero elements.
Events with one jet and one $b$-jet will have the following indices with non-zero values: 2, 3, 12, 13, and 18.
For events with MET and a single electron, the values will be non-zero at indices 1, 5, 10, and 15.
For two-muon events, the non-zero values will be at indices 4, 14, 22, and 37.

On the physics side, the RMM-C46 representation attempts to retain the full RMM information for leading objects in transverse energy, while compressing the information from softer objects by aggregating their variables into a small set of features. By contrast, the original RMM preserves the full information for each object up to the chosen fixed multiplicity ($N=10$), as discussed in Sect.~\ref{sec:RMM_definition}. As a result, RMM-C46 can be effective for event topologies dominated by a few high-momentum jets or particles, but it may be less suitable for high-multiplicity events where the detailed properties of low transverse-energy objects are important for ML training.

Finally, in studies where the signal region is defined in terms of invariant masses (as in \cite{ATLAS:2023ixc}), which must be modeled with a smoothly defined background function and therefore should not be biased by the ML technique, RMM-C46 enables the explicit removal of such masses from the ML inputs. This feature is similar to the original RMM. For example, if the signal region is defined as the dijet invariant-mass distribution, $m_{j,j}$ values at the index 32 must be excluded from the ML input.

\section{Performance of the compressed RMM representations}
\label{sec:performance}

\subsection{Comparative performance in Supervised Machine learning}

To assess the discriminating power of RMM-C46 relative to the full sparse RMM, we trained a lightweight fully-connected neural network on a binary classification task.
The benchmark uses simulated $X\!\rightarrow\! SH$ events with $m_X=1500~\text{GeV}$
as signal and $t\bar{t}$ dileptonic events as background, corresponding to the
same physics setup used in the RMM literature.

A simple multilayer perceptron (MLP) with two hidden layers (ReLU activations,
dropout regularization, and a sigmoid output) was trained on each of the four
feature formats:
\begin{itemize}
    \item the full RMM (excluding the first four metadata values),
    \item the 46-dimensional RMM-C46 vector (Frobenious norms),

\end{itemize}
All inputs were normalized and the same architecture and hyperparameters were
applied across formats to enable a direct comparison.

The resulting area under the ROC curve (AUC) values are:

\[
\text{AUC}(\text{Full RMM}) = 0.998, \text{AUC}(\text{RMM-C46}) = 0.999.
\]

The full RMM, containing thousands of correlated entries, unsurprisingly
provides excellent separation between the signal and background.  Remarkably, both compact representations achieve nearly the same performance while reducing the dimensionality by orders of magnitude.

The RMM-C46 format separates the physics channels within each block-mass-like, rapidity-like, and transverse-energy structures-yielding a richer and more faithful compression of the original matrix.

In summary, RMM-C46 provides a high-fidelity, physics-aware compression that remains extremely lightweight for machine learning, yet preserves-and in this test even slightly exceeds the discriminating power of the full RMM.

\subsection{Performance of RMM-C46 with Unsupervised Anomaly Detection with Autoencoders}
\label{sec:AE_unsup}

To assess the information content of the various compact RMM representations,
we performed an unsupervised anomaly–detection study using a simple
fully–connected autoencoder (AE).  
The AE is trained \emph{only on Standard Model background events} and therefore
learns to reproduce the kinematic correlations typical of QCD and
$t\bar{t}$ production.  
Any event whose topology deviates from this learned manifold-such as a heavy
resonance decay-is expected to yield a larger reconstruction error.

\paragraph{Training procedure.}
For each representation (RMM and RMM-C46), we proceed as follows:
\begin{enumerate}

\item \textbf{Background-only training.}  
The $t\bar{t}$ background sample is split into
training, validation, and test partitions.  
Only the background training subset is used to optimize the autoencoder parameters.

\item \textbf{Standardization.}  
Each representation is standardized using statistics computed exclusively
from the background training set:
\[
X_{\text{scaled}} = \frac{X - \mu_{\text{bkg}}}{\sigma_{\text{bkg}}}\, .
\]

\item \textbf{Autoencoder architecture.}  
The network consists of a bottleneck architecture,
\[
d_{\text{in}} \rightarrow 64 \rightarrow 32 \rightarrow 16 \rightarrow
32 \rightarrow 64 \rightarrow d_{\text{in}},
\]
optimized using Adam with early stopping based on validation loss.

\item \textbf{Reconstruction loss.}  
For any event $x$, the anomaly score is the mean-squared reconstruction error,
\[
\mathcal{L}(x) = \frac{1}{d_{\text{in}}}
\left\| x - \hat{x} \right\|_2^2,
\]
where $\hat{x}$ is the AE output.

\item \textbf{Signal evaluation.}  
After training on background, the AE is applied to
independent $t\bar{t}$ test events and to
$X\!\rightarrow\!SH$ signal events at $m_X = 1500\,$GeV.  
Signal events typically yield larger losses because their
topologies lie outside the learned background manifold.
\end{enumerate}

\paragraph{AUC computation.}
The reconstruction-loss distributions for background and signal
are treated as classifier scores and used to build an ROC curve:
background is assigned label~0, signal label~1.
The area under the ROC curve (AUC) quantifies how well the AE
separates signal from background in an \emph{entirely unsupervised} manner.

\paragraph{Results and interpretation.}

The anomaly detection performance, quantified using the area under the ROC curve (AUC), yields

\[
\text{AUC}(\text{Full RMM}) = 0.9865,\qquad
\text{AUC}(\text{RMM-C46}) = 0.9995.
\]

\begin{figure}[H]
    \centering
    \includegraphics[width=1.0\textwidth]{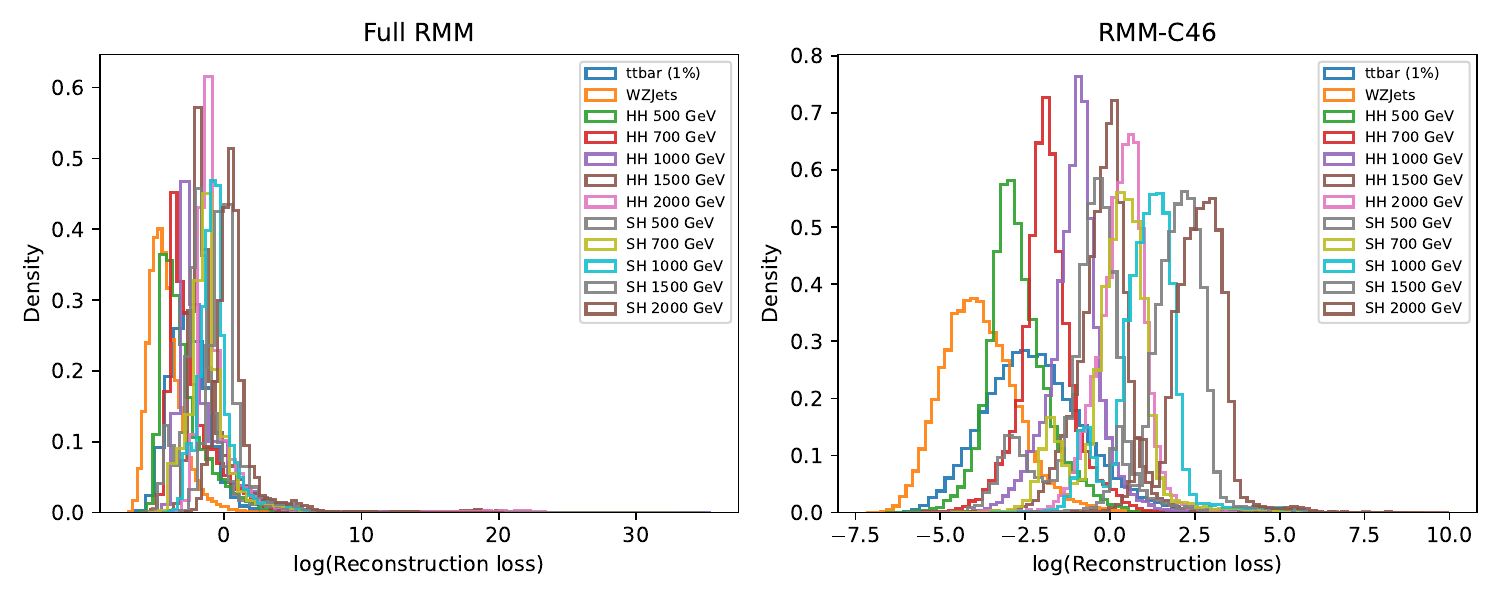}
    \caption{
    Comparison of the loss distributions obtained using the RMM and RMM-C46 inputs. The autoencoders have identical architectures. The loss distributions are shown for different MC processes.}
    \label{fig:ROC_loss_AE_multi_1}
\end{figure}

\begin{figure}[H]
    \centering
    \includegraphics[width=1.0\textwidth]{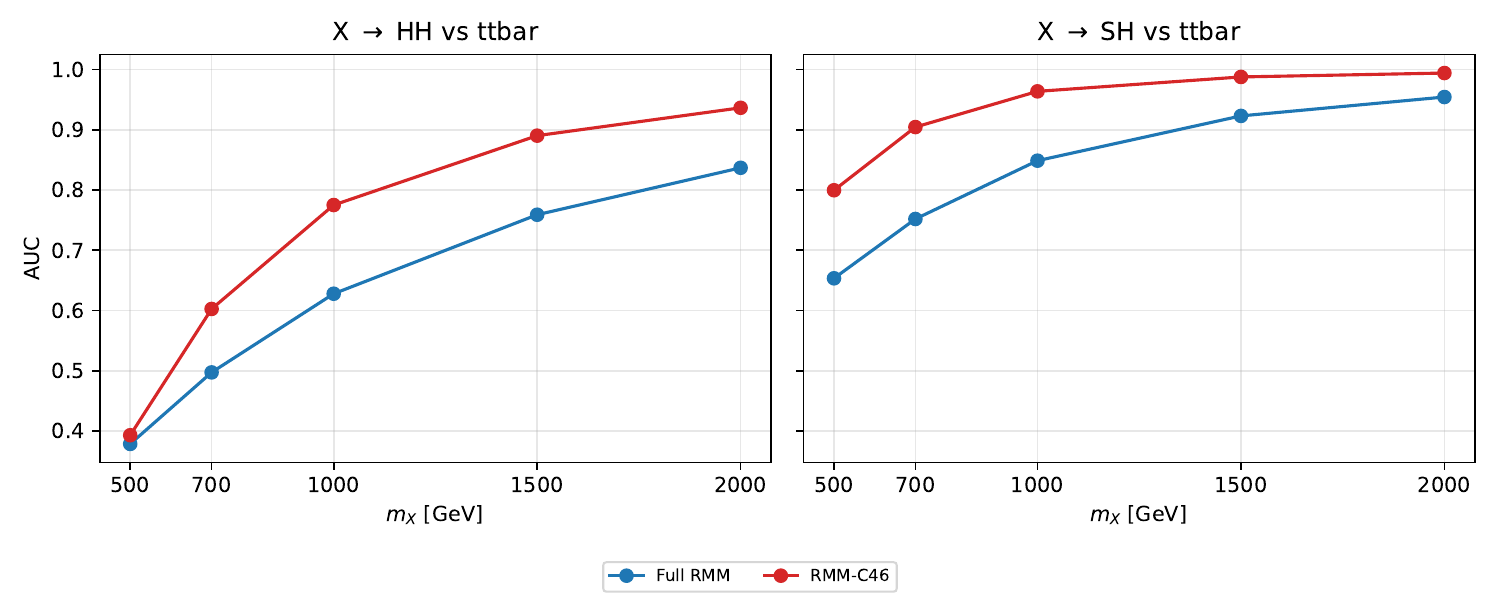}
    \caption{
    Comparison of AUC curves for the signal Monte Carlo vs $t\bar{t}$ background, after applying the selection cut on the reconstruction losses. They are computed using identical autoencoder architectures, with inputs provided either as RMM or RMM-C46.}
    \label{fig:ROC_loss_AE_multi_2}
\end{figure}

Figures~\ref{fig:ROC_loss_AE_multi_1},~\ref{fig:ROC_loss_AE_multi_2} illustrate the corresponding reconstruction loss distributions and AUC trends obtained using identical autoencoder architectures with inputs provided either as the full RMM and its compact C46 representation. The loss distributions are shown for multiple Monte Carlo (MC) processes, highlighting the separation between signal and background in the learned latent space. The corresponding AUC curves, computed for signal MC against the $t\bar{t}$ background after applying selection cuts on the reconstruction losses, further demonstrate the improved discrimination achieved with the compressed representation.

These results indicate that the compact C46 representation preserves nearly all of the intrinsic discriminative manifold structure present in the full high-dimensional RMM. Notably, the C46 input outperforms the full RMM in this unsupervised setting, suggesting that the removal of sparsely populated, noisy, or redundant regions of the original matrix enhances the autoencoder’s ability to learn a faithful representation of the background manifold. In particular, the Frobenius-aggregated C46 (frob), constructed using physically motivated zone-wise norms, achieves the highest level of signal–background separation. This demonstrates that block-aggregated, physics-driven features provide a cleaner and more learnable low-dimensional manifold than the raw. 

Thus, the compact RMM-C46 input space for ML offers a powerful, interpretable, and computationally efficient alternative to the full RMM for unsupervised anomaly detection at the LHC.

\begin{figure}[H]
    \centering
    \includegraphics[width=1.0\textwidth]{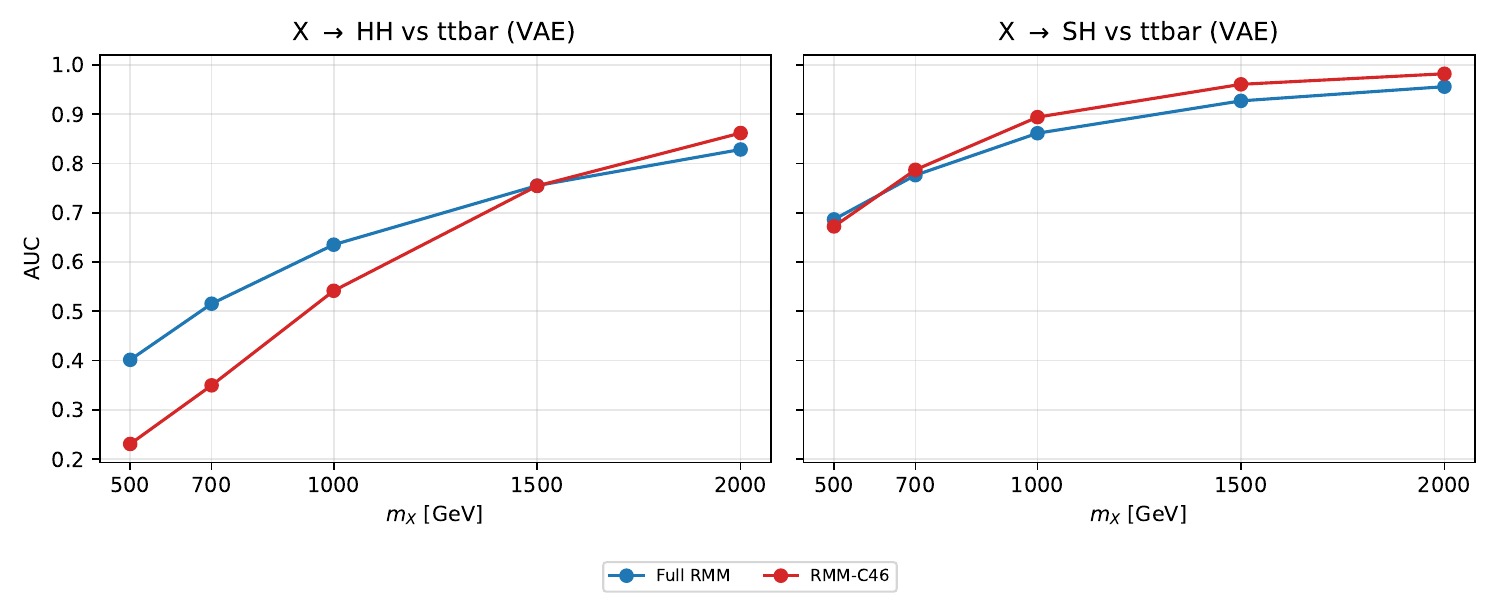}
    \caption{
    Comparison of AUC curves for the signal MC  vs $t\bar{t}$ background, after applying the selection cut on the reconstruction losses. They are computed using identical variational autoencoder (VAE) with inputs provided either as RMM or RMM-C46.}
    \label{fig:ROC_loss_VAE_multi}
\end{figure}

The VAE used in this study adopts the same layer configuration as the vanilla autoencoder. The key distinction lies in the treatment of the latent representation. Instead of mapping the 32-dimensional hidden layer to a single deterministic 16-dimensional bottleneck, the VAE introduces two parallel linear projections that parameterize a latent Gaussian distribution through its mean $\mu$ and log-variance $\log\sigma^{2}$. A latent vector is then sampled using the reparameterization trick, ensuring differentiability during backpropagation. The decoder receives this sampled latent vector and mirrors the encoder structure to reconstruct the input. Consequently, while the VAE shares the same layer sizes as the vanilla autoencoder, it incorporates a probabilistic latent space and a Kullback--Leibler regularization term, yielding a more expressive generative model for anomaly detection tasks.

Figure~\ref{fig:ROC_loss_VAE_multi} shows the AUC performance of the VAE-based reconstruction–loss anomaly detector for HH and SH signals across different resonance masses $m_X$. A clear improvement with increasing mass is observed in all cases; however, the compact RMM-C46 representation consistently matches or surpasses the Full RMM at intermediate and high masses. Notably, RMM-C46 delivers comparable discrimination at low masses while achieving superior AUC values as $m_X$ increases, demonstrating that a physics-guided compact representation can outperform a much larger high-dimensional input. For both HH and SH channels, RMM-C46 achieves excellent separation (AUC $\gtrsim 0.9$) at the highest mass points.

A comparison between the vanilla autoencoder and the VAE indicates that both architectures successfully capture the increasing separation power between signal and background as the resonance mass $m_X$ grows, though their detailed behavior differs. The AE generally exhibits stronger discrimination at lower masses, particularly for the compact C46 representation, as seen from its steeper rise in AUC in Fig.~\ref{fig:ROC_loss_AE_multi_2}. In contrast, the VAE yields a smoother  and more uniform scaling with $m_X$, with the compact RMM-C46 representation consistently matching or exceeding the Full RMM performance at intermediate and high masses (Fig.~\ref{fig:ROC_loss_VAE_multi}). This behavior reflects the influence of the probabilistic latent space: while the AE captures fine-grained deterministic structure, the VAE provides a more regularized reconstruction model that becomes increasingly effective for heavier signals whose kinematic patterns deviate more substantially from the background manifold. Overall, both models perform extremely well at large $m_X$, with RMM-C46 emerging as a highly competitive and efficient compact representation across both architectures.

\subsection{Other compact representations}

We investigated a few other compact representations, summarizing each interaction class (type--type or MET--type) and producing a compact set of ``correlation-strength'' observables, where each 
variable encodes the aggregate coupling between two object classes, but does not preserve a clean factorization of the underlying RMM structure.

In comparison, the RMM-C46 representation offers a more granular and physically transparent decomposition.  The full RMM is partitioned into 46 non-overlapping zones corresponding to distinct physical quantities:  
the MET scalar ($1$), transverse-energy diagonals ($5$),  
transverse-mass sectors ($5$), longitudinal/Lorentz sectors ($5$),  
rapidity-difference blocks ($15$), and invariant-mass blocks ($15$).  
In the Frobenius variant (RMM-C46-frob), each variable corresponds to the 
Frobenius norm of \emph{exactly} one such RMM zone.  
This yields a structured physics bookkeeping in which, for example, 
``total jet--jet invariant mass activity'' and 
``total muon transverse-mass strength'' are represented by distinct, 
well-defined components. In this sense, RMM-C46 provides a more interpretable and physically factorized summary of the event kinematics, compared to other compact representations we investigated.


\subsection{Correlation structure of the RMM-C46 representation}

\begin{figure}[H]
    \centering
    \includegraphics[width=1.0\textwidth]{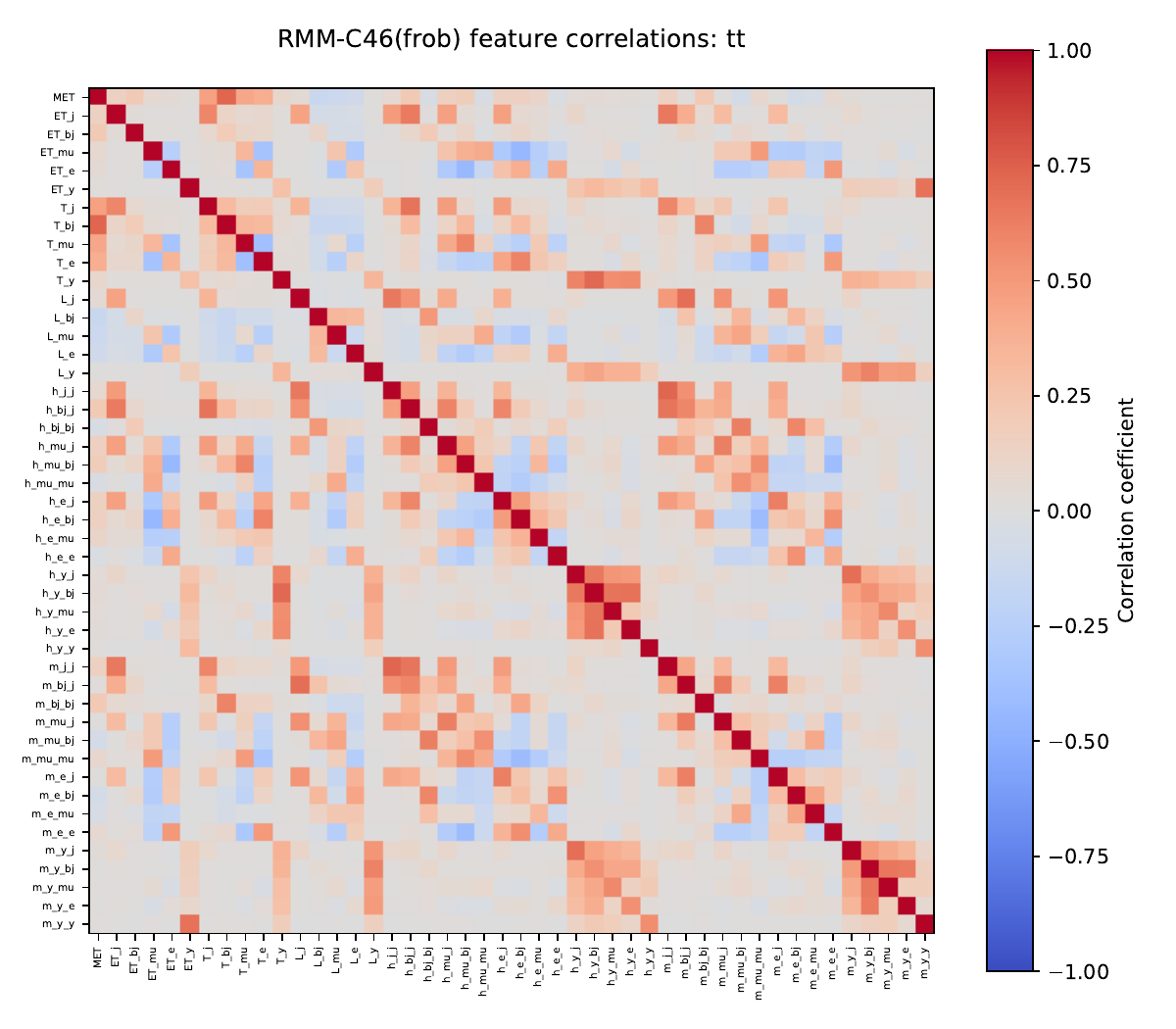}
    \caption{
    Correlation in the RMM-C46 event representation of $\mathrm{t\bar t}$ samples}
    \label{fig:corr_tt}
\end{figure}

Figures~\ref{fig:corr_tt} presents a correlation matrix computed using all available events in $\mathrm{t\bar t}$ sample. For the full-sample matrices, each entry represents the Pearson (linear) correlation between two RMM-C46 features evaluated across all events in the given sample, treating each event as one observation in the statistical ensemble.

For every pair of features, the matrix entry represents the Pearson correlation coefficient, defined as
\[
\rho(x,y)
=
\frac{
\sum_{i=1}^{N} (x_i - \bar{x}) (y_i - \bar{y})
}{
\sqrt{
\sum_{i=1}^{N} (x_i - \bar{x})^{2}
}
\;
\sqrt{
\sum_{i=1}^{N} (y_i - \bar{y})^{2}
}
},
\]
where $\bar{x}$ and $\bar{y}$ denote the sample means of the two features,
\[
\bar{x} = \frac{1}{N} \sum_{i=1}^{N} x_i,
\qquad
\bar{y} = \frac{1}{N} \sum_{i=1}^{N} y_i.
\]

The coefficient takes values in the range $-1 \le \rho(x,y) \le 1$ and can be interpreted as follows:
\begin{itemize}[nosep]
    \item \textbf{+1 (dark red):} perfect positive linear correlation - the two
          features always increase together;
    \item \textbf{--1 (dark blue):} perfect negative linear correlation - one
          feature increases when the other decreases;
    \item \textbf{0 (light grey):} no linear correlation - the features vary
          independently.
\end{itemize}
Entries that are undefined (e.g.\ due to zero variance) or exactly zero are shown as white, indicating ``no information'' or completely empty correlations.

Across all processes, similar trends were observed, and the strong diagonal dominance and the presence of clear block-like structures demonstrate that the 46 compressed variables retain the physics-driven factorization inherent in the rapidity-mass matrix.  Sectors corresponding to the same object class (jet–jet, $b$-jet–$b$-jet, muon–muon, etc.) form coherent positively correlated regions, reflecting that their Frobenius norms summarize related sets of invariant-mass and rapidity-difference cells. Cross-type interactions exhibit weaker and more heterogeneous correlations, as expected from the more stochastic population of mixed-object blocks in the original RMM.

Among the background processes, the $\mathrm{ttbar}$ and $\mathrm{WZJets}$
correlation matrices display broadly similar patterns, with moderate
correlations among the jet and $b$-jet sectors and nearly uncorrelated lepton and photon sectors.  
In contrast, the heavy-resonance signals ($HH$ and $SH$) exhibit noticeably stronger correlations within jet and $b$-jet zones, as well as in MET--jet interactions.  

These patterns arise from the higher-$E_T$, multi-jet final states typical of TeV-scale scalar decays, which populate a larger fraction of the RMM invariant-mass
and rapidity blocks with non-zero entries, leading to stronger and more uniform feature correlations.
 
The $SH$ sample shows the most pronounced correlation structure, with dense, coherent positive blocks linking jets, $b$-jets, and MET, characteristic of the boosted multi-Higgs cascade topology.

We emphasize that these correlation matrices are used purely as diagnostic and visualization tools; the downstream machine-learning models operate directly on the event-level RMM-C46 feature vectors.


\section{Prospects for Quantum and Hybrid Implementations}
\label{sec:quantum_prospects}

The compact RMM representation of collision events developed in this work, the so-called  RMM-C46,  provides a natural foundation for next-generation ML studies, including resource-efficient classical training and 
deployable quantum or hybrid quantum–classical pipelines.  
Its central advantage lies in transforming the high-dimensional, heavily correlated 
structure of LHC events into a small collection of interpretable, physics-grounded 
features that preserve event topology while dramatically reducing computational cost.

\subsection{Motivation for compressed collider representations}

The original RMM input space contains more than a thousand entries per event (for example, 1287 elements for ten jets, five leptons, 
and five photons).  
Although this richness captures fine-grained multi-object correlations, it also 
creates substantial challenges for both classical and quantum ML.  
Training large neural networks on such high-dimensional inputs is slow and 
memory-intensive, and the abundance of sparsity and noise can lead to unstable 
optimization or overfitting.  
Moreover, quantum devices with limited qubit counts cannot accommodate such 
large feature spaces without aggressive preprocessing.

The RMM-C46 representation mitigates all of these issues by compressing the full matrix into 46 physics-defined observables that summarize energy flow, transverse 
components, rapidities, and invariant-mass structures.  
Constructed through simple summations or Frobenius norms over well-defined zones, 
these variables preserve the physical semantics of each interaction channel while 
eliminating redundant, noisy, or sparsely populated regions of the original matrix.
The compressed representation reduces kinematic information for low-momentum objects. Therefore, we expect it to perform best for events containing several hard, well-separated jets, leptons, and photons.

\subsection{Advantages for classical ML training}

Reducing the input dimensionality from thousands of raw RMM entries to a structured 
set of only a few dozen features substantially accelerates and stabilizes classical 
ML workflows.  
Networks operating on RMM-C46 input can be trained faster, require fewer hyperparameter 
iterations, and tend to generalize better due to the removal of physically irrelevant 
noise.  
Interpretability is preserved because each feature corresponds to a specific, 
well-defined physics block, allowing downstream analyses-such as supervised 
classification or unsupervised anomaly detection-to operate in a well-conditioned 
feature space with reduced covariance structure.  
Empirically, we find that RMM-C46 matches or exceeds the discriminating performance 
of the full RMM in several ML tasks, despite using only a few percent of the original 
dimensionality.  
This compact structure also facilitates rapid prototyping of new model architectures 
without the engineering overhead typically associated with thousand-dimensional inputs.

\subsection{Bridging to Quantum and Hybrid Quantum-Classical Machine Learning}

Near-term noisy intermediate-scale quantum (NISQ) devices provide only $\mathcal{O}(10)$ high-fidelity qubits with limited entangling depth, rendering direct encoding of high-dimensional collider data impractical. The RMM-C46 representation offers a hardware-compatible alternative: its forty-six normalized, physics-structured features can be efficiently mapped onto angle or amplitude encodings using approximately $\mathcal{O}(10)$ qubits, depending on the chosen protocol. The bounded and smooth nature of the Frobenius-compressed variables mitigates dynamic-range instabilities during quantum state preparation, while their organization into mass-like, rapidity-like, and transverse-energy sectors enables symmetry-aware quantum feature maps aligned with collider kinematics. By compressing the rapidity–mass matrix into a compact and physically factorized representation, RMM-C46 provides a realistic and interpretable interface between collider data and emerging quantum machine-learning architectures.

\subsection{Outlook for the HL-LHC era}

The high luminosity of future LHC runs will generate unprecedented amounts of 
collision data, magnifying the need for compact, interpretable, and computationally 
efficient representations.  
RMM-C46 offers a standardized low-dimensional interface for both classical and 
quantum ML, providing a physics-grounded feature set compatible with 
cross-experiment comparisons and scalable to large datasets.  
Its low dimensionality aligns naturally with the resource limitations of NISQ 
hardware, while its interpretability facilitates robust physics validation, even as 
ML algorithms become increasingly sophisticated.  
As quantum technologies improve in qubit count and coherence time, RMM-C46 provides 
a realistic pathway toward quantum-enhanced workflows for collider-physics analyses.

\section{Conclusion}

In summary, the RMM-C46 representation of collision events serves as a powerful bridge between 
traditional high-dimensional descriptions of final states produced in particle collisions and the resource constraints of 
modern ML systems.  
Its interpretability, computational efficiency, and intrinsic compatibility with 
emerging quantum technologies make it a promising foundation for next-generation 
classical and hybrid quantum–classical analyses at the LHC and future colliders.  
By distilling complex event topologies into a compact and physics-informed format, 
RMM-C46 enables scalable learning pipelines and opens the door to practical 
quantum machine-learning applications in high-energy physics. The source code of RMM-C46 is publicly available from the GIT repository~\cite{c46git}.




\section*{Acknowledgments}
WI is supported by DE-SC0017647.SC is supported by UChicago Argonne, LLC, operator of Argonne National Laboratory, under Contract No. DE-AC02-06CH11357 with the U.S. DOE. Argonne’s work is supported by the DOE Office of High Energy Physics under the same contract. We also acknowledge the computing resources provided by the Laboratory Computing Resource Center at Argonne National Laboratory.


\bibliographystyle{JHEP}
\bibliography{references}



\clearpage
\appendix






\end{document}